\documentstyle[11pt,subeqn]{article}
\begin{document}
\begin{titlepage}

\vspace{1.5in}
\begin{center}
{\bf \Large The Age of the Universe, the Hubble Constant and
QSOs in a Locally Inhomogeneous Universe\\}
\vspace{.75in}
{\bf J. W. Moffat $^{\dag}$ \& D. C. Tatarski $^{\ddag}$\\}
{\bf Department of Physics\\}
{\bf University of Toronto\\}
{\bf Toronto, Ontario M5S 1A7, Canada\\}
\vspace{.5in}

Talk given at the XIV Moriond Workshop on Particle Astrophysics, Atomic
Physics and Gravitaton, Villars sur Ollon, Switzerland, January 22-29, 1994.

\vspace{.5in}
{\bf\large Abstract}
\end{center}
A local void in the globally Friedmann-Robertson-Walker cosmological
model with the critical density ($\Omega_{0}=1$) is studied. The
inhomogeneity is described using a Lema\^{\i}tre-Tolman-Bondi solution
for a spherically symmetric distribution of matter. The scale of the central
underdense region is $\sim 150$ Mpc. We investigate the effects this has
on the cosmological time scale, the measurement of the Hubble constant
and the redshift--luminosity distance for moderately and very distant objects
($z \sim 0.1$ and more). The results indicate that if we happened to live in
such a void, but insisted on interpreting cosmological observations through
the FRW model, we could go wrong in a few instances. For example, the
Hubble constant measurement could give results depending on the
separation of the source and the observer, the quasars could be younger
than we think and also less distant (less energetic).

\vspace{.5in}
\dag: moffat@medb.physics.utoronto.ca

\ddag: tatarski@medb.physics.utoronto.ca

\vspace{.25in}
{\bf \large UTPT-94-09, Revised: May 1994.}

\end{titlepage}

\section{Introduction}

It seems, particularly after the introduction of the inflationary paradigm
\cite{guth}, that the isotropic and homogeneous Friedmann-Robertson-
Walker (FRW)  cosmological models are best suited for the description of
the global structure and the evolution of the universe. However, a similar
statement is not necessarily true when cosmologically moderate scales are
thought of.

There exists direct observational evidence in favour of the isotropy of the
observed universe, namely, the COBE data confirming a high degree of
isotropy of the cosmic microwave background radiation (CMBR)
\cite{cobe,COBE}. However, there is no observationally based reason
supporting the assumption of homogeneity. On the contrary, there seems to
exist observational evidence in favour of larger and larger structures
\cite{lstr}.

Recent work on modelling voids in the expanding universe \cite{bc} has
shown that it is possible to construct an asymptotically FRW universe
containing an expanding spherical void of the Lema\^{\i}tre-Tolman-Bondi
type \footnote{A cosmological solution spherically symmetric about one
point was first proposed by Lema\^{\i}tre \cite{lem}.  However, it is usually
called the Tolman--Bondi solution \cite{tol,bon}.}. Moreover, the void can
be a region of under-density rather than vacuum.

We think that there exists sufficient observational evidence (briefly
discussed later in this paper) to support a conjecture that we may live in a
relatively large underdense region embedded in a globally FRW universe.
Exploring physical properties of such a model is the aim of the present
paper.

In the following section, we briefly discuss the LTB model. Section
\ref{void}
consists of a brief discussion of the observational background and the
simple toy model of a local void presented here. The closing section
contains a description of our results of numerical calculations as well as
conclusions.

Throughout this paper we use units in which $G=c=1$, unless stated
otherwise.

\section{The Model} \label{model}

First, for the sake of notational clarity, let us recall the
FRW line element:
\begin{equation} \label{frwlinel}
 ds^{2}=dt^{2}-{a}^{2}(t)\left[ \frac{dr^{2}}{1-kr^{2}} +         
 r^{2}d\Omega^{2} \right],
\end{equation}
with $d\Omega^{2}={d\theta}^{2}+{\mbox{sin}}^{2}\theta{d\phi}^{2}$.

Now, let us consider a Lema\^{\i}tre-Tolman--Bondi \cite{tol,bon}
 model for a spherically symmetric inhomogeneous universe filled with
dust. The line element in comoving coordinates can be written as: 
\begin{equation} \label{linel}
 ds^{2}=dt^{2}-R^{\prime 2}(t,r)f^{-2}dr^{2}-R^{2}(t,r)d\Omega^{2},
\end{equation}
where {\em f} is an arbitrary function of {\em r} only, and the
field equations demand that {\em R(t,r)} satisfies:
\begin{equation} \label{F}
 2R\dot{R}^{2}+2R(1-f^{2})=F(r),
\end{equation}
with {\em F} being an arbitrary function of class $C^{2}$,
$\dot{R}=\partial R / \partial t$ and $R^{\prime}=\partial R /
\partial r$. We have three distinct solutions depending on whether
$f^{2}<1$, $=1$, $>1$ and they correspond to elliptic (closed),
parabolic (flat) and hyperbolic (open) cases, respectively.

The proper density can be expressed as:
\begin{equation} \label{dens}
 \rho=\frac{F^{\prime}}{16\pi R^{\prime} R^{2}}.
\end{equation}

Whatever the curvature, the total mass within comoving radius $r$ is:
\begin{equation} \label{mass}
M(r)= \frac{1}{4} \int_{0}^{r} dr f^{-1} F^{\prime} =4 \pi
\int_{0}^{r} dr \rho f^{-1} R^{\prime} R^{2},
\end{equation}
so that
\[
 M^{\prime} (r) = \frac{dM}{dr} = 4 \pi \rho f^{-1} R^{\prime}
R^{2}.
\]
Also for $\rho > 0$ everywhere we have $F^{\prime} >0$ and
$R^{\prime} >0$ so that in the non-singular part of the model $R>0$
except for $r=0$ and $F(r)$ is non-negative and monotonically
increasing for $r \geq 0$. This could be used to define the new
radial coordinate $\bar{r}^{3}=M(r)$ and find the parametric solutions
for the rate of expansion.

In the flat (parabolic) case $f^{2}=1$, we have
\begin{equation}
R = \frac{1}{2}{\left(9F\right)}^{1/3}{\left(t+\beta\right)}^{2/3},
\end{equation}
with $\beta(r)$ being an arbitrary function of class $C^{2}$ for
all $r$. After the change of coordinates $R(t,\bar{r}) = \bar{r}
{\left(t+\beta(\bar{r})\right)}^{2/3}$, the metric becomes:
\begin{equation} \label{metric}
 ds^{2}=dt^{2}-{\left(t+ \beta \right)}^{4/3} \left( Y^{2}        
  dr^{2}+r^{2}d\Omega^{2} \right),
\end{equation}
where
\begin{equation}
 Y= 1 + \frac{2 r {\beta}^{\prime}}{3 \left(t+ \beta \right)},
\end{equation}
and from (\ref{dens}) the density is given by
\begin{equation} \label{densb}
 \rho = \frac{1}{6 \pi {\left(t+ \beta \right)}^{2} Y} .
\end{equation}
Clearly, we have that ($t \rightarrow \infty$) the model tends to the flat
Einstein--de Sitter case.

For the closed and open cases the parametric solutions for the rate
of expansion can be written as \cite{bonnor}:
\begin{subequations} \label{parneg}
\begin{equation}
 R=\frac{1}{4} F {\left( 1 - f^{2} \right)}^{-1} \left[1-\cos(v)
\right] ,\quad    f^{2}<1 ,
\end{equation}
\begin{equation}
 t+\beta= \frac{1}{4} F {\left( 1-f^{2} \right)}^{-3/2}\left[ v-
\sin(v) \right]   ,\quad f^{2}<1 ,
\end{equation}
\end{subequations}
and
\begin{subequations} \label{parpos}
\begin{equation} \label{parposR}
 R= \frac{1}{4} F {\left( f^{2}-1\right)}^{-1} \left[\cosh(v)-1
\right] ,\quad   f^{2}>1 ,
\end{equation}
\begin{equation} \label{parpost}
 t+\beta= \frac{1}{4} F {\left( f^{2}-1\right)}^{-3/2} \left[
\sinh(v)-v \right]   ,\quad f^{2}>1 ,
\end{equation}
\end{subequations}
with $\beta(r)$ being again a function of integration of class
$C^{2}$ and $v$ the parameter.

The flat case ($f^{2}=1$) has been rather extensively studied elsewhere
\cite{motat}. The model depends on one arbitrary function $\beta (r)$ and
could be specified by assuming the density on some space-like
hypersurface, say $t=t_{0}$.

The cases of interest to us, (\ref{parneg}) and (\ref{parpos}), correspond to
closed and open models, respectively.

Before we proceed (in the next section) to discuss the observational grounds
for modelling a local void, we need to amplify the discussion of the LTB
model by introducing basic features of the propagation of light in our
model. The high degree of isotropy of the microwave background forces us
to the conclusion that we must be located very close to the spatial centre of
the void. In our discussion, for the sake of simplicity, we place an observer
at the centre ($t_{Ob}=t_{0} , r_{Ob}=0$).

The luminosity distance between an observer at the origin of our
coordinate system ($t_{0},0$) and the source at
($t_{e},r_{e},\theta_{e},\phi_{e}$) is \cite{bon}:
\begin{equation} \label{lumdis}
 d_{L}={\left(\frac{{\cal L}}{4\pi{\cal F}}\right)}^{1/2}=R(t_{e},r_{e}){\left[1
 + z(t_{e},r_{e}) \right]}^2,
\end{equation}
where ${\cal L}$ is the absolute luminosity of the source (the
energy emitted per unit time in the source's rest frame), ${\cal
F}$ is the measured flux (the energy per unit time per unit area as
measured by the observer) and $z(t_{e},r_{e})$ is the redshift
(blueshift) for a light ray emitted at ($t_{e},r_{e}$) and observed
at ($t_{0},0$).

The light ray travelling inwards to the centre satisfies:
\[
 ds^{2}=dt^{2}-R^{\prime 2}(t,r)f^{-2}dr^{2}=0 ,\quad d\theta=d\phi=0,
\]
and thus
\begin{equation} \label{zero}
 \frac{dt}{dr}=-R^{\prime}(t,r)/f(r) .
\end{equation}

Without getting into a detailed discussion, which can be found in
\cite{bon,motat}, let us state that if the equation of the light ray travelling
along the light cone is:
\begin{equation} \label{rays}
 t = T(r) ,
\end{equation}
using (\ref{zero}) we get the equation of a ray along the path:
\begin{equation} \label{lightcone}
 \frac{dT(r)}{dr}=-\frac{R^{\prime}}{f} [T(r),r] ,
\end{equation}
where 
\[
 {\dot{R}}^{\prime}[T(r),r] = 
 {\left. \frac{{\partial}^{2}R}{\partial t \partial r} \right     
 |}_{r,T(r)}   ={\left. \frac{\partial R^{\prime}}{\partial t}    
 \right |}_{r,T(r)} .
\]
The equation for the redshift considered as a function of $r$ along the light
cone is:
\begin{equation} \label{zred}
 \frac{dz}{dr}=(1+z){\dot{R}}^{\prime}[T(r),r] ,
\end{equation}
and the shift $z_{1}$ for a light ray travelling from ($t_{1},r_{1}$) to
($t_{0},0$) is:
\begin{equation} \label{grco}
  \log (1+z_{1})  =  - \log (1-a_{1}) - \int_{0}^{r_{1}} dr                    
  \frac{M^{\prime}(r)}{r(1-a_{1})} ,
\end{equation}
where
\[
a_{1}(r)=\dot{R}[T(r),r] ,
\]
and, in obtaining the second equation, we used (\ref{dens}) and
(\ref{mass}). Thus we have two contributions to the redshift. The
cosmological redshift due to expansion, described by the first term
with $a_{1}=\dot{R}$, and the gravitational shift due to the difference
between the potential energy per unit mass at the source and at the
observer. Obviously, in the homogeneous case ($M^{\prime}(r)=0$)
there is no gravitational shift.

\section{The modelling of the local void} \label{void}

If we restrict ourselves to spatial scales that have been well probed
observationally, i.e. up to a few hundred Mpc, the most striking feature of
the luminous matter distribution is the existence of large voids surrounded
by sheet-like structures containing galaxies (e.g. \cite{grwall}). The
surveys \cite{grwall},\cite{iras} give a typical size of the voids
of the order 50--60 $h^{-1}$ Mpc. There has also been some evidence
\cite{broadhurst} --with less certainty-- for the existence of
larger underdense regions with characteristic sizes of about 130
$h^{-1}$ Mpc. Also, dynamical estimates of the FRW density parameter
${\Omega}_{0}$ give very different results on different scales. The
observations of galactic halos on scales less than about 10 to 30 Mpc
typically give (see e.g. \cite{sancisi}) ${\Omega}_{10-30} \simeq 0.2 \pm
0.1$. On the other hand, smoothing the observations over larger scales
($>20 \mbox{ Mpc}$, say $\sim 100 \mbox{ Mpc }$) indicates (e.g.
\cite{iras}) the existence of a less clustered component with a contribution
exceeding 0.2, and perhaps as high as ${\Omega}_{\sim 100} \simeq 0.8
\pm 0.2$.

At the same time, the large scale galaxy surveys (some of the recent
literature is given in \cite{surveys}) firmly indicate a considerable excess in
the number--magnitude counts for faint galaxies relative to predictions of
homogeneous, ``no-evolution'' models. This excess could be the result of  a
non-standard galactic evolution or could be caused by rather exotic FRW
cosmology (i.e. the deceleration parameter $q_{0} \ll 0.5$ or a non-zero
cosmological constant $\Lambda$). However, it can also be treated as an
observational indication of a very large (on the scale of the redshift $z \sim
0.5$) void. A model of a local void with a density distribution based on the
faint galaxies number counts is presently being studied \cite{nmotat}.

In the model presented here, we confine ourselves to the simple density
distributions. We study two cases: a void with the central density equal to
that of an FRW model with the density parameter $\Omega_{0}=0.2$,
asymptotically approaching the FRW model with $\Omega_{0}=1$, and a
very similar void ``distorted'' on an intermediate scale by a peak ($\Omega
\stackrel{<}{\sim} 1$) in the density distribution. The two distributions are:
\begin{subequations} \label{densdistr}
\begin{equation} \label{densdistrv}
\Omega_{v}(r)=\Omega_{min}+\left(\Omega_{max}-\Omega_{min} \right)
\left[1-{\left(\frac{r}{L}\right)}^2 \frac{exp(r/L)}{{\left[exp(r/L)-1
\right]}^2}\right],
\end{equation}
and
\begin{equation} \label{densdistrvp}
\Omega_{vp}(r)=\Omega_{min}+\left(\Omega_{max}-\Omega_{min} \right)
\left[1-{\left(\frac{r}{L}\right)}^2 \frac{exp(r/L)}{{\left[exp(r/L)-1
\right]}^2}\right] + {\left(\frac{r}{l}\right)}^2
 e^{-(r/l)^2}.
\end{equation}
\end{subequations}
In the numerical calculations presented in the next section we used the
values $\Omega_{min}=0.2$, $\Omega_{max}=1$ and $L=l=30
\mbox{Mpc}$. This assures that the void converges satisfactorily fast to the
outside critical FRW universe ($\Omega \simeq 0.86$ for $r \simeq 150
\mbox{Mpc}$ and $\Omega \simeq 0.95$ for $r \simeq 200
\mbox{Mpc}$) and that the intermediate peak is observationally acceptable
($\Omega \simeq 0.64$ for $r \simeq 33 \mbox{Mpc}$ ).

\section{The results and discussion} \label{resdis}

In general, an LTB model depends on three arbitrary functions, see section
\ref{model}, $F(r), \beta(r)$ and $f(r)$. Since $F(r)$ can be interpreted as
twice the effective gravitational mass within comoving radius $r$
(\cite{bon}), then, in accordance with the discussion following (\ref{mass}),
assuming its form is equivalent to a coordinate choice. In our calculations
we used $F(r)=4r^3$. The second function, $\beta(r)$, sets the initial
singularity (``big bang'') hypersurface of the model. Since we want the
outside region in our toy model to be fully equivalent to the critical FRW
universe, we set $\beta(r)=0$, thereby assuming a universally simultaneous
big bang. We also set the time coordinate of constant time hypersurface
``now'' so that it is equal to the age of the universe $t_{0}$ in the FRW
model with $\Omega_{0}=1$. In doing so, we give up a very important
feature of an LTB model: an extra (with respect to FRW) degree of freedom
that would allow the age of the universe to be different from FRW or even
position dependent. The third (``curvature'') function, $f(r)$, is an unknown
to be solved for in our calculations. From the work done in \cite{bc}, we
conclude that the LTB case to be used in modelling an underdense
comoving void in an FRW universe is the hyperbolic ($f^2>1$) one.

In a manner similar to that employed in \cite{motat}, we assume that since
all cosmological observations are necessarily done by detecting some form
of electromagnetic radiation, the solution should progress along the light
cone. The final set of equations we solve consists of the equations
(\ref{lightcone}), (\ref{zred}) and the equation describing the density
distribution (\ref{dens}) through either of the relations (\ref{densdistr}) taken
along the light cone, e.g.:
\begin{equation} \label{findens}
\rho[T(r),r]=\frac{F^{\prime}(r)}{16\pi R^{\prime}[T(r),r] R^{2}[T(r),r]}=
\Omega_{v}(r).
\end{equation}
Since $T(r)$ (the time of emission $t_{e}$ of a light ray observed at $r=0$ at
$t_{0}$) is now given by (\ref{parpost}), the functions to be solved for are
$f(r), z(r)$ and $v(r)$, where the parameter $v$ becomes the function of
position. The initial conditions for the integration have to be set at $r \neq
0$, since the analytic expressions (\ref{parpos}) are singular at $r=0$,
where
$f^2=1$ (we have a flat $\Omega_{0}=0.2$ FRW universe there). We
assume that for the initial radius $r_i \ll 1$ (we use dimensionless radius
and time in the calculations) corresponding $z_i$ and $t_i$ are given by
their standard FRW values. Then $z(r_i), v(r_i)$ and $f(r_i)$ can be obtained
from (\ref{parpost}), (\ref{lightcone}) and (\ref{zred}).

Once the equations have been numerically integrated we use (\ref{parpost})
to obtain $t_{e}$ for a given $r_{e}$. The luminosity distance $d_{L}$
corresponding to this event is obtained with the use of (\ref{lumdis}) and
$z(t)$ (useful in studying the cosmological time scale) is given by the
parametric relation $[ T(r), z(r) ]$.

\begin{figure}[ht]
\vspace{3.2in} \relax \noindent \relax
\includegraphics{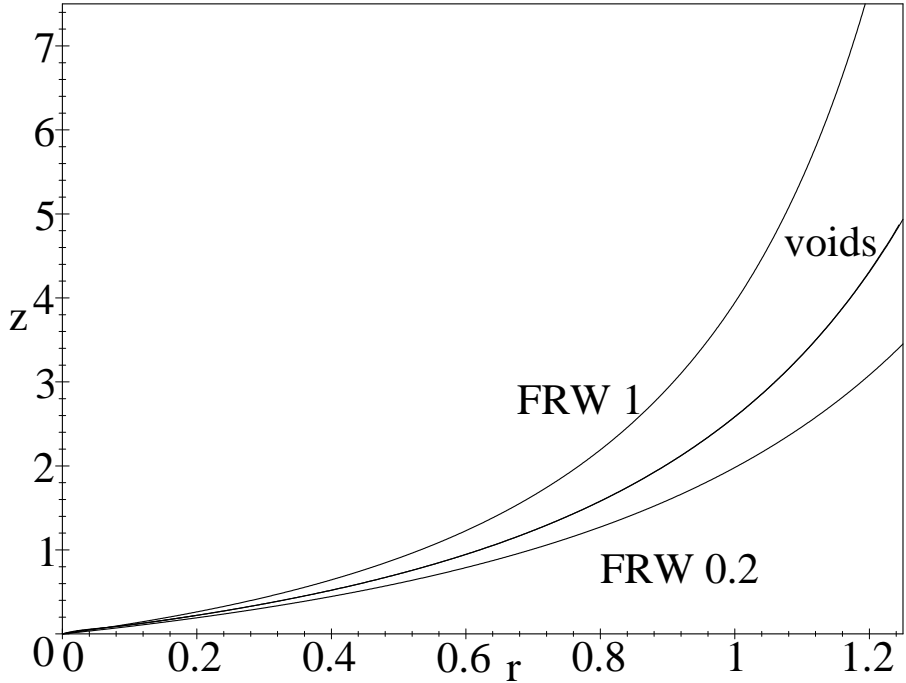}
\caption{{\tenrm\baselineskip=10pt The redshift $z$ as a function of
comoving radius $r$. The FRW results for $\Omega=1$ and
$\Omega=0.2$ are denoted by ``FRW 1'' and ``FRW 0.2'', respectively.
The results for the LTB modelled voids are denoted ``voids''. \label{f1}}}
\end{figure}

The results of our numerical calculations are as follows. Figure \ref{f1}
depicts $z(r)$, where $r$ is the dimensionless comoving radius used in the
calculations. The coordinate distance has no direct physical relevance, but
our units here are such that $r=1$ corresponds to $2997.95 h^{-1}$ Mpc,
where $h$ is the usual coefficient in the observationally determined value
of the Hubble constant: $H_{0}=100 h \mbox{ km } {\mbox{s}}^{-1}
{\mbox{Mpc}^{-1}}$.

\begin{figure}[ht]
\vspace{3.2in} \relax \noindent \relax
\includegraphics{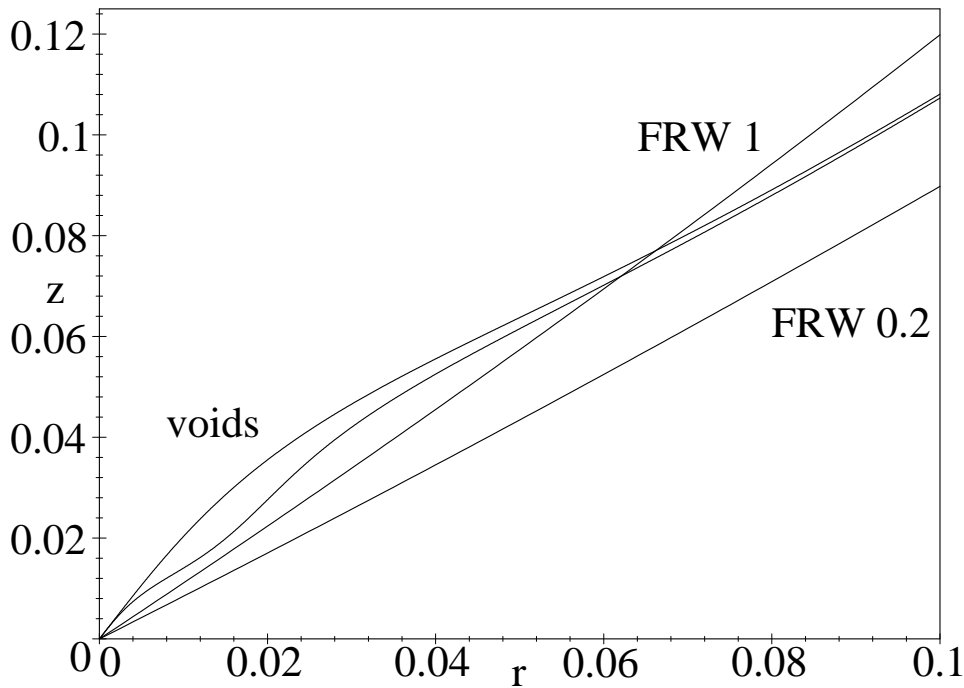}
\caption{{\tenrm\baselineskip=10pt The redshift $z$ as a function of
comoving radius $r$. Smaller scales. Notation as in Fig. 1. \label{f2}}}
\end{figure}

The departure of $z(r)$ from its FRW behaviour does not seem to be
dramatic. In fact, on the distance scale used in Figure \ref{f1} it is hardly
noticeable. More details, on a smaller spatial scale, can be seen in Figure
\ref{f2}. The redshift $z$, after being influenced noticeably by rapid
changes in the density $\rho$ on smaller scales, asymptotically tends to a
limit that could, in accordance with FRW interpretation, correspond to the
universe with the density parameter in the range $\Omega \in (0.2;1)$.
\footnote{This situation does not change considerably if we model the
central void in a similar manner but with respect to the observable $z$ as a
variable \cite{nmotat}.} The increase in $z$ on intermediate scales is clearly
induced by the additional gravitational shift caused by the mass distribution
of the void. The large scale behaviour is controlled by our assumption of
the equality $t_0=t_{0 FRW}$.

However, one should not forget that the comoving distance is {\em not} an
observable, whereas the luminosity distance $d_L$ is. In principle, provided
we know its absolute luminosity $\cal{L}$, we can establish the luminosity
distance, defined by $4 \pi {d_L}^2=\cal{L} / \cal{F}$, by measuring the
energy flux $\cal{F}$ of an observed object (for a discussion of usual
caveats associated with so-called ``standard candles'' see e.g.
\cite{cosmoloc}).

Due to the lack of space we do not present the $z(t)$ relation here. There
is, again, some small scale ($t_0-t \ll 1$) divergence from the FRW
behaviour, but for early cosmological times ($t \ll 1$) the relation tends to
the critical FRW ($\Omega=1$) one. This is in accordance with our
assumption of a simultaneous big bang, $\beta(r)=0$, and with our setting
the age of the universe to be equal to that of the critical FRW case. Objects
with redshifts of order a few are younger than their FRW counterparts,
but not significantly.

\begin{figure}[ht]
\vspace{3.2in} \relax \noindent \relax
\includegraphics{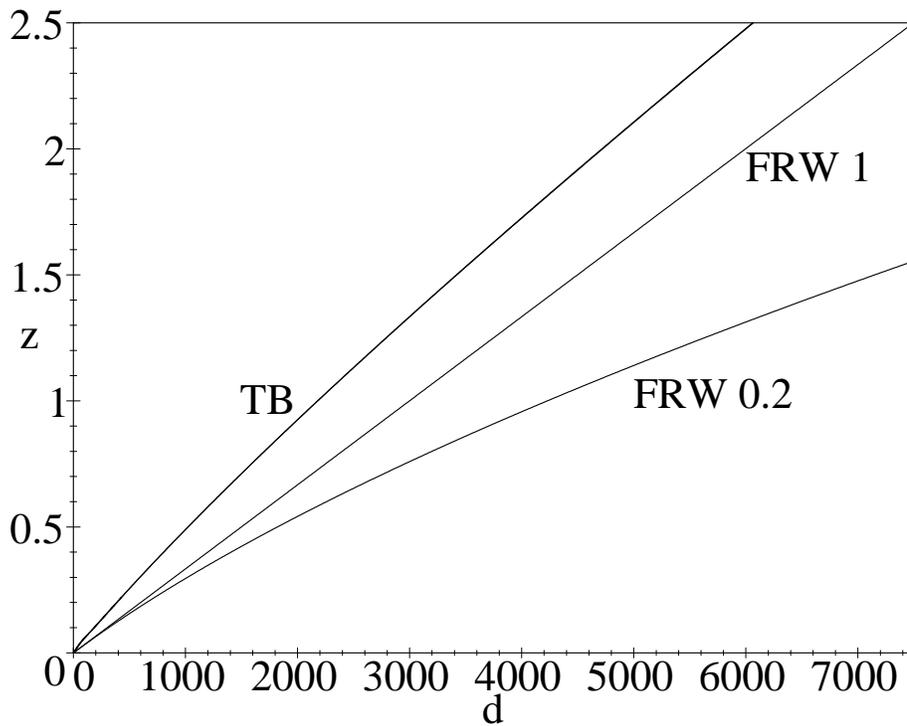}
\caption{{\tenrm\baselineskip=10pt The redshift $z$ as a function of
the luminosity distance $d_L$ (in Mpc). \label{f3}}}
\end{figure}

\begin{figure}[hb]
\vspace{3.2in} \relax \noindent \relax
\includegraphics{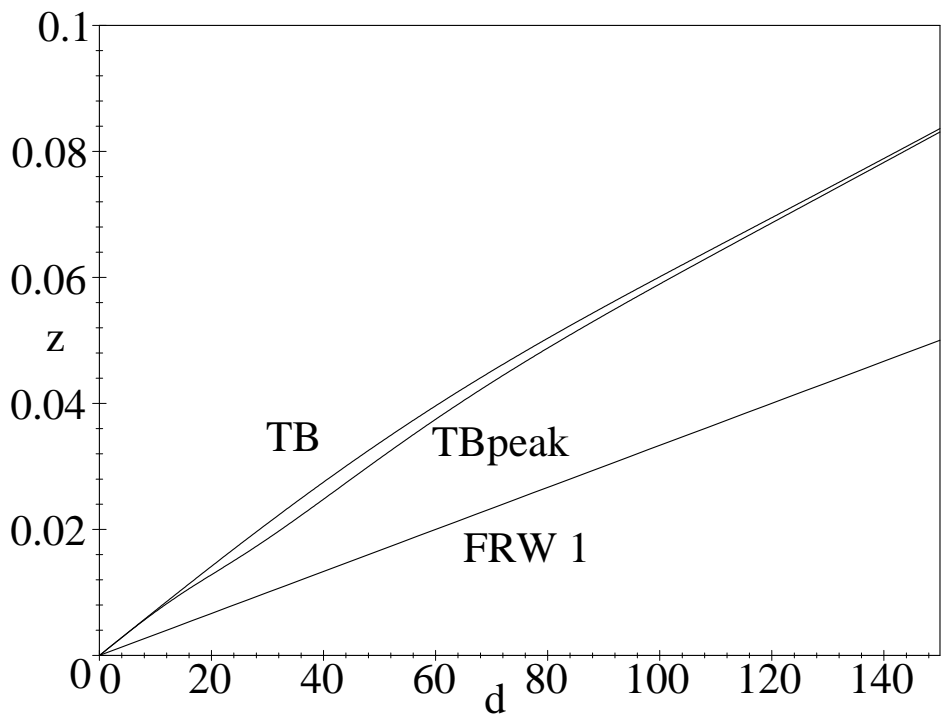}
\caption{{\tenrm\baselineskip=10pt The redshift $z$ as a function of
the luminosity distance $d_L$ (in Mpc). Smaller scales. \label{f4}}}
\end{figure}

Figures \ref{f3} and \ref{f4} present the redshift--luminosity distance
relation on cosmologically large (Figure \ref{f3}) and intermediate (Figure
\ref{f4}) scales. Now the departure from the FRW behaviour becomes
apparent. This comes as no surprise if one recalls the formulae for $d_L$ in
both LTB (\ref{lumdis}) and FRW:
\begin{equation}
{d_L}^2=a^2(t_0) {r_e}^2 {(1+z_e)}^2 .
\end{equation}
The important question is whether our results contradict the linearity of the 
Hubble relation $z=H d$, well established on small scales.
\pagebreak

\nopagebreak

\vspace{0.5in}
\nopagebreak

\pagebreak

Due to our choice of the cosmological time scale and FRW embedding, the
asymptotic ($z \rightarrow \infty$) behaviour of $z(d_L)$ is that of the
critical ($\Omega=1$) FRW case. On intermediate scales, however, objects
of comparable redshifts are located at smaller luminosity distances. The
ratio $d_{LTB}/d_{FRW}$ is $\approx 0.7$ ($\approx 0.5$) for $z \approx
1$, $\approx 0.8$ ($\approx 0.4$) for $z \approx 2.5$ and $\approx 0.9$
($\approx 0.36$) for $z\approx4.5$, where the most distant quasars are
observed. (Values in parentheses correspond to the FRW $\Omega=0.2$
case.) Since the absolute luminosity $\cal{L}$ of the source scales as the
square of $d_L$ this reduces the energy output of QSOs (up to an order of
magnitude). Also, the angular diameter distance $d_A = D/ \delta =
{(1+z)}^{-2} d_L$ now gives a smaller proper distance $D$ across the
source for the same observed angular diameter $\delta$. This helps resolve
problems with seemingly acausal signals (correlations in luminosity bursts)
observed across some quasars.

At the same time, inspection of Figure \ref{f4} shows that on small scales a
very nearly linear (in fact, observationally indistinguishable from linear)
``Hubble diagram'' is obtained. However, a different value for the Hubble
parameter (constant) is inferred (position, or rather $d_L$, dependent on
larger scales), if we insist on interpreting the results of cosmological
observations through an FRW model.

To explore this possibility let us recall that in FRW cosmology the exact
result for the Hubble relation ($z$ versus $d_L$) in the matter dominated
universe is \cite{cosmoloc}:
\begin{equation} \label{hubblefrw}
H_0 d_L ={q_0}^{-2} \left[ zq_0 + \left(q_0-1\right)\left(\sqrt{2zq_0+1} -1
\right) \right],
\end{equation}
where $q_0 \equiv -\ddot{a}(t_0)/a(t_0){H_0}^2$ is the deceleration
parameter.

Let us assume that we live in a local LTB void and the $z$ vs. $d_L$
relation differs from the FRW one as described in this paper, but we are
biased by our theoretical prejudice and interpret cosmological observations
through the FRW model.

\begin{figure}[ht]
\vspace{3.2in} \relax \noindent \relax
\includegraphics{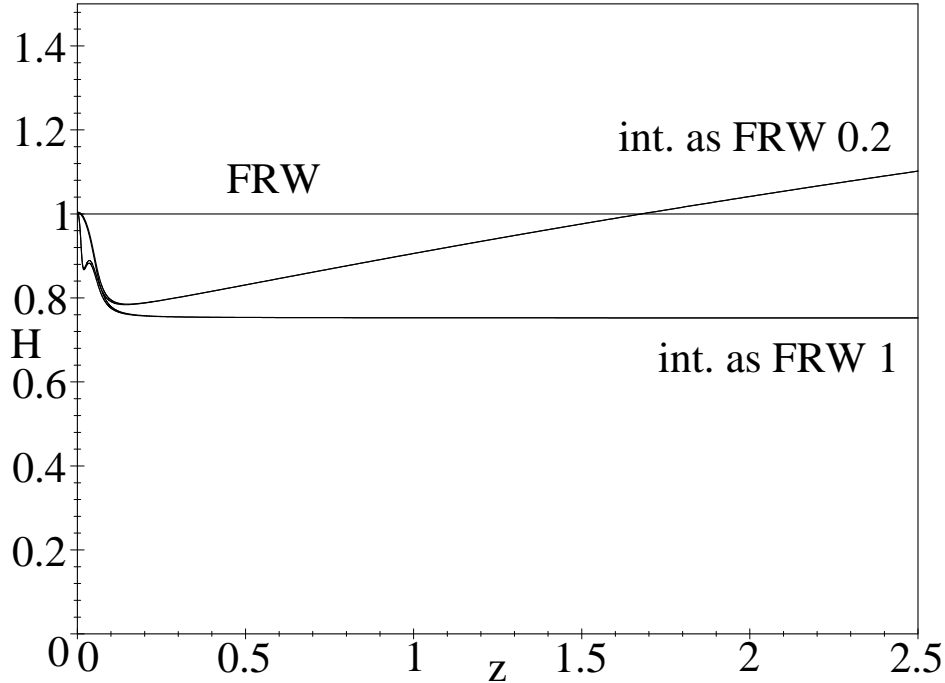}
\caption{{\tenrm\baselineskip=10pt The ``observed'' Hubble constant
$H_0$ (in units of the local measurement) as a function of the redshift $z$.
The results interpreted through FRW $\Omega=0.2$ and $\Omega=1$
models,
respectively, are denoted ``int. as FRW 0.2'' and ``int. as FRW 1''.
\label{f5}}}
\end{figure}

On cosmologically small distances we measure the same value of $H_0$
independently of the model (we call this value ``the local measurement'').
This stems from the fact that, due to our assumptions, very close to the
centre ($r \ll 1$) the model is well approximated by the FRW universe with
$\Omega=0.2$. Obviously, if the universe were LTB rather than FRW, then
the Hubble parameter based on the observed (LTB) values of $z$ and
$d_L$, but inferred through an FRW relation (\ref{hubblefrw}), would be
position (redshift) dependent as shown in Figure \ref{f5}.

The values of $H_0$ reported to date span the range $40$ to $100 \mbox{
km } {\mbox{s}}^{-1} {\mbox{Mpc}^{-1}}$ (with standard errors quoted
frequently as $10 \mbox{ km } {\mbox{s}}^{-1} {\mbox{Mpc}^{-1}}$ or
less!). Inhomogeneities similar to the LTB void presented here might provide
an explanation for this.

The LTB void, such as the one presented here, decreases its density contrast
(the depth of the void with respect to the FRW background) when evolved
back in time \cite{bonnor,motat}. At early times it is almost homogenized
(at $t/t_0 \simeq 10^{-5}$ we have $|{\rho}_{LTB}(r)/{\rho}_{FRW}-1| <
10^{-6}$). This corresponds to a universe which at the beginning is very
similar to the FRW one, but different at late stages.

In this manner, while retaining all accomplishments of the FRW cosmology
in dealing with epochs preceding the matter dominated era, we can gain
new freedom in modelling the more recent universe. We can solve the age
of the universe problem (by assuming $\beta(r) = const \neq 0$),
provide the excess power observed on scales of 5---10,000 km s$^{-1}$ in
modelling structure formation (see \cite{motat}), alleviate a few old
problems associated with quasars (their age, luminosity and size) and
provide an explanation for the wide range of reported values of the Hubble
constant.

\vspace{0.5cm}
\noindent {\bf Acknowledgement}

The authors thank Ken-Ichi Nakao for helpful discussions and suggestions.

\end{document}